\documentclass[twocolumn,superscriptaddress,prl,epsfigs]{revtex4} 
\usepackage{graphicx} 
\usepackage{amssymb} 
\usepackage{amsmath} 
\usepackage{appendix} 
\usepackage{color} 
 
\newcommand{\addMR}[1]{\textcolor{red}{#1}}

\def\blank#1{} 
\newcommand{\be}{\begin{equation}} 
\newcommand{\ee}{\end{equation}} 
\newcommand{\bea}{\begin{eqnarray}} 
\newcommand{\eea}{\end{eqnarray}}

\newcommand{\lp}{\left(} 
\newcommand{\rp}{\right)}

\newcommand{\mpar}[1]{\marginpar{\small\it #1}}

\begin{document} 
 
\title{  
Chirality-Assisted Electronic Cloaking in Bilayer Graphene Nanostructures} 
\author{Nan Gu} 
\affiliation{Department of Physics, Massachusetts Institute of Technology, Cambridge, Massachusetts 02139, USA} 
\author{Mark Rudner} 
\affiliation{Department of Physics, Harvard University, Cambridge, Massachusetts 02138, USA} 
\author{Leonid Levitov} 
\affiliation{Department of Physics, Massachusetts Institute of Technology, Cambridge, Massachusetts 02139, USA} 
\begin{abstract} 
We show that the strong coupling of pseudospin orientation and charge carrier motion in bilayer graphene has a drastic effect on transport properties of ballistic p-n-p junctions. 
Electronic states with zero momentum parallel to the barrier are confined under it for one pseudospin orientation,
whereas states with the opposite pseudospin tunnel through the junction totally uninfluenced by the presence of confined states. We demonstrate that the junction acts as a cloak for confined states, making them nearly invisible to electrons in the outer regions over a range of incidence angles. This behavior is manifested in the two-terminal conductance as transmission resonances with non-Lorentzian, singular peak shapes. The response of these phenomena to a weak magnetic field or electric-field-induced interlayer gap can serve as an experimental fingerprint of electronic cloaking. 
\end{abstract} 
 
\maketitle


Charge carriers in graphene behave like relativistic particles\cite{Beenakker2008}.
Some of the most intriguing aspects of carrier dynamics in this material arise due to chirality, i.e.~the strong coupling of pseudospin and orbital degrees of freedom\cite{Peres2010}. 
Analogs of relativistic electron effects such as Klein tunneling\cite{Katsnelson, Cheianov, Pereira2010}, and optical phenomena such as negative refraction\cite{Cheianov2007}, Fabry-P\'erot resonances\cite{Shytov,Barbier2009,Darancet2009} and the Goos-H\"anschen effect\cite{Beenakker2009,Sharma}, provide a platform for understanding transport in graphene nanostructures\cite{Stander,Gorbachev,Young}. 
Chirality was also proposed as a vehicle for coupling the orbital motion of carriers to the inner valley degrees of freedom of graphene\cite{Rycerz2007,Garcia-Pomar2008}.

\begin{figure} 
\includegraphics[width=3.2in]{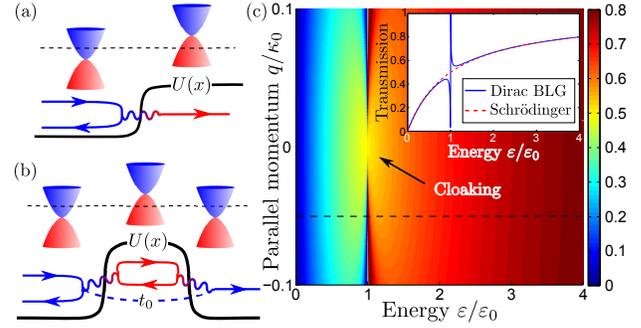} 
\caption[]{Cloaking of transmission resonances in ballistic p-n-p junctions due to decoupling of states with orthogonal pseudospins. 
For a potential step (a), the pseudospin-reversing coupling 
(wavy line) vanishes at normal incidence.
This leads to formation of confined states under a barrier (b), 
seen as narrow resonances in transmission.
The cloak effect 
illustrated in (c) for a delta function barrier,
having one confined state which produces a transmission resonance $\epsilon=\epsilon_0$ (see Eq.(\ref{eq:delta_function}) and accompanying text). 
The resonance is narrowed and dimmed at near-normal incidence angles (small $q$). As seen in the inset, showing trace along dashed line,
transmission at all $\epsilon\ne\epsilon_0$ is uninfluenced by the confined state
(the cloak effect).
}
\vspace{-8mm}
\label{fig1} 
\end{figure}

Here we describe new effects in bilayer graphene (BLG) that have no direct analogs in optics or in single-layer graphene. These effects, which arise due to the chiral nature of carriers, have dramatic consequences for transport through potential barriers. 
At normal incidence, 
chirality mismatch leads to complete decoupling of states in regions of opposite polarity
(see Fig.\ref{fig1}a,b). 
Electrons of one chirality are confined within the barrier, despite the presence of a continuum of 
available states 
outside the barrier. 
Conversely, electrons in the outer region 
scatter and tunnel through 
the barrier 
as if no localized 
states were 
available on their way. 
Thus the barrier acts as a {\it cloak} for confined states, rendering them invisible via both transmission and reflection.
This effect, which is unique to BLG, leads to a number of intriguing and potentially useful properties of BLG nanostructures,
such as tunable confinement and coupling to individual states.

\begin{figure} 
\includegraphics[width=3.25in]{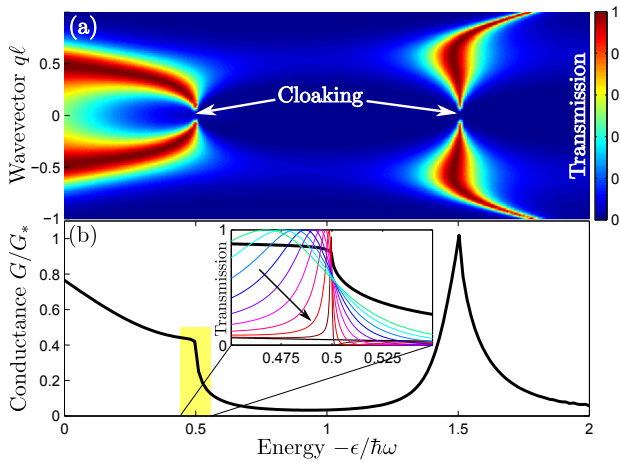} 
\caption[]{The transmission probability (a) and the conductance (b) for parabolic barrier $U(x) = -\frac12 m \omega^2 x^2$. 
Shown are 
resonances corresponding to the first two harmonic oscillator bound states, $\epsilon = -\frac12 \hbar \omega$ and $-\frac32 \hbar \omega$.
At normal incidence, $q=0$, these confined states are cloaked by the barrier due to chirality mismatch with the continuum states outside.
 The conductance resonances feature the characteristic square-root singularity $\delta G \propto -\sqrt{|\delta \epsilon|}$. Units:
$G_* = G_0 W/\ell$, where $W$ is the sample width and $\ell = \sqrt{\hbar/2m\omega}$.  A zoom-in shows the first resonance overlaid with 10 transmission profiles, with $q\ell$ changing from $0.18$ to $0$ as indicated by an arrow.
} 
\vspace{-6mm}
\label{fig2} 
\end{figure}

\begin{figure}[t] 
\includegraphics[width=3.25in]{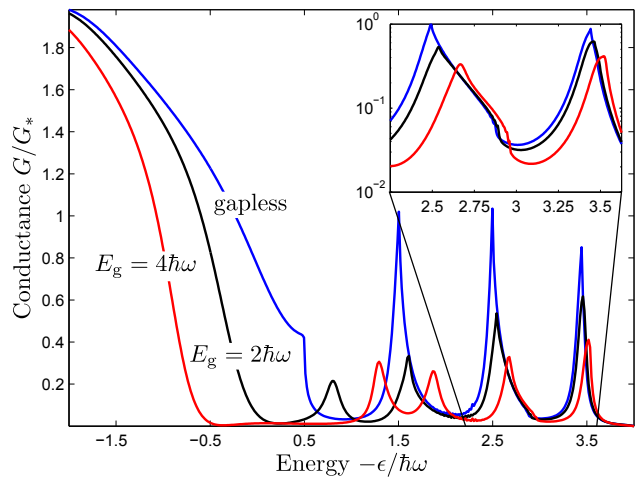} 
\caption[]{The effect of interlayer gap on ballistic conductance for a parabolic barrier, $U(x) = -\frac12 m \omega^2x^2$. Cloaking manifests itself in abnormally sharp resonances positioned at $\epsilon = -\hbar \omega(n + \frac12)$, with $n = 0, 1, 2, 3$ for a gapless system. The sharp cusps observed in the gapless case, Eq.(\ref{eq:G}), are 
rounded by a nonzero gap. Results for gap values $E_{\rm g}=0$, $2\hbar \omega$, and $4\hbar \omega$ are shown.  
Zoom-in on the 
resonances at $\epsilon = -\frac52\hbar \omega, -\frac72\hbar \omega$ (inset) 
indicates 
that an overall suppression of conductance, estimated from the max/min ratio, has little effect on the fringe contrast. 
}
\vspace{-6mm}
\label{fig:G3} 
\end{figure}

For oblique incidence, pseudospin decoupling is imperfect,
producing transmission resonances associated with confined states. However, these resonances are very narrow at near-normal incidence angles, making 
the outer states `blind' to the confined states at almost all energies (see Figs. \ref{fig1}c and \ref{fig2}). 
This yields a high-fidelity cloak effect for all near-normal incidence angles. 
The effect is sensitive to external magnetic field applied perpendicular to BLG plane: the Lorentz force curves particle trajectories, making the orbits normally incident on the barrier exit at oblique angles. We show that the narrowing of transmission resonances at normal incidence is manifested in sharp non-Lorentzian peaks in the two-terminal conductance with square-root singularities at the tips (see Figs.\,\ref{fig2} and \ref{fig:G3}). 
This singular behavior, along with its suppression in magnetic field, can serve as an experimental signature of the cloak effect. 

It is instructive to compare the electron cloak realized in BLG with optical cloaks of invisibility\cite{Pendry2006,Leonhardt2006}, which employ 
refractive metamaterial shells 
to guide plane waves around an object. In contrast, in our approach, the probing wave is transmitted directly through the region containing the cloaked states, made invisible by decoupling of opposite chiralities. Crucially, cloaked states can either hold particles or be empty. Since electron interactions in BLG are nearly pseudospin-blind\cite{Peres2010}, filling of confined states does not affect the decoupling of opposite chiralities which is responsible for the cloak effect. For similar reasons, states with different spin and valley polarization can be cloaked simultaneously.

We also note that scattering on barriers in BLG was studied in Refs.\cite{Katsnelson,Barbier2009}, where Fabry-P\'erot resonances at oblique incidence were found. Analytic results for transmission through a square barrier at normal incidence were obtained\cite{Katsnelson}, however the cloak effect and its relation to confined states was not elucidated.

In experiments, several factors may modify the picture of chirality-induced cloaking of confined states. First, disorder in the barrier region, which breaks conservation of the parallel momentum $q$, contributes additional broadening of the resonances, in particular near $q=0$. However, recent advances in experimental techniques, such as using suspended graphene samples~\cite{Feldman}, and deposition on hexagonal boron nitride~\cite{Dean}, help to dramatically enhance the mobility of BLG samples and minimize the influence of disorder. 
Second, the top and back gates used to create p-n-p junctions in graphene devices naturally create a potential difference between the two layers, thus opening a gap in the BLG spectrum~\cite{Trickey, McCann}. The effect of gap opening will be discussed below.
 
Electronic transport in BLG in the presence of a potential barrier $U(x)$ is governed by the Hamiltonian\cite{McCann} 
\be \label{eq:H} H = \left(\begin{array}{cc} 
U(x)  &          -\frac{\hbar^2}{2m}(\frac{d}{dx} + q)^2 \\ 
-\frac{\hbar^2}{2m}(\frac{d}{dx} - q)^2 &   U(x) 
\end{array}\right),
\ee 
where the barrier is spatially uniform in the $y$-direction, and $q$ is the conserved wavevector  
component parallel to it ($m \approx 0.04 m_e$ is the BLG band mass). 
The cloak effect is complete 
for normally incident electronic states, for which $q=0$. 
The problem can be conveniently analyzed in the eigenbasis of $\sigma_x$, $\sigma_x |\pm \rangle = \pm |\pm \rangle$. 
In this basis the pseudospinor Schr\"{o}dinger equation 
decouples into two scalar Schr\"{o}dinger equations:
\be \label{eq:coupleEq0}
\left(-\frac{\hbar^2}{2m}\frac{d^2}{dx^2} \pm 
(U(x) - \epsilon)\right)\psi_\pm(x) = 0.
\ee 
For the special case of a potential step (see Fig.\ref{fig1}a),  
and for energies between the values of $U(x)$ far to the left and far to the right, $U_{\rm n}<\epsilon<U_{\rm p}$, 
plane wave states on one side are coupled to evanescent states on the opposite side.  
Thus, despite the availability of free carrier states with energy $\epsilon$ on either side of the step,
the transmission is completely blocked due to chirality mismatch.

Extending this argument, we show that chirality leads to the confinement of electronic states under potential barriers in BLG. For a general form of potential barrier $U(x)$, as shown in Fig.\ref{fig1}b, the two sign choices in Eq.(\ref{eq:coupleEq0}) lead to two qualitatively distinct types of states with zero momentum parallel to the barrier ($q=0$). The solutions for $\psi_+(x)$ describe continuum states that live outside the barrier region, while the solutions for $\psi_-(x)$ describe bound states confined in the inverted potential $-U(x)$. Because these confined states completely decouple from the continuum states at $q=0$, they do not show up in the normal incidence transmission (c.f.~Ref.\cite{Katsnelson}). 
 
Crucially, transmission through the barrier in BLG is controlled by different mechanisms at normal incidence and at oblique incidence. At normal incidence, since pseudospin is conserved, the only available mode of transmission is direct tunneling via the evanescent wave that extends through the barrier. The character of transmission changes completely for oblique incidence. Extending Eq.(\ref{eq:coupleEq0}) to nonzero $q$, we obtain two coupled equations: 
\be \label{eq:coupleEq} 
\left[-\frac{d^2}{dx^2} - q^2 \pm \frac{2m}{\hbar^2}(U(x)-\epsilon)\right]\psi_\pm(x) =  
2q\frac{d\psi_\mp(x)}{dx}. 
\ee 
Due to the $\psi_+$/$\psi_-$ mixing, the confined states acquire finite coupling to the continuum, turning into transmission resonances. However, since the width of these resonances vanishes as $q^2$ at near-normal incidence, the cloak effect persists in a finite range of incidence angles with small $q$.

We highlight various aspects of the cloaking behavior by analyzing two models, 
a narrow barrier and a wide barrier, modeled by delta function and inverted parabolic potentials.
In the first case, $U(x)=\lambda\delta(x)$, $\lambda>0$, at zero $q$ we get a confined state 
$\psi_-(x)\propto e^{-\kappa_0|x|}$, with $\kappa_0=\lambda m/\hbar^2$ 
and energy $\epsilon_0=\lambda^2m/2\hbar^2$.
Scattering states for a delta function can be found exactly, by extending the standard approach 
to account for evanescent states which appear in the free particle BLG problem. 

We nondimensionalize Eq.(\ref{eq:coupleEq0}) via $\epsilon\to (\hbar^2/2m)\epsilon$, $\lambda\to (\hbar^2/2m)\lambda$, and write the free particle wavefunction as
\be
\psi_\pm(x>0)=a_\pm e^{-\kappa x}+b_\pm e^{ik x}+c_\pm e^{-ik x}
,\quad
\epsilon>q^2
,
\ee
with $k=\sqrt{\epsilon+q^2}$, $\kappa=\sqrt{\epsilon-q^2}$. For a potential with inversion symmetry, $U(x)=U(-x)$, there are two types of symmetric solutions of Eq.(\ref{eq:coupleEq0}): those corresponding to $\psi_+(x)$  even and $\psi_-(x)$ odd, and vice versa.  
Taking $\psi_+(-x)=\pm \psi_+(x)$, $\psi_-(-x)=\mp \psi_-(x)$ in each case, and performing standard matching for the values and derivatives at $x=0$, we find a relation between in and out states of the form $b_+^{(1,2)}=z_{(1,2)}c_+^{(1,2)}$, where the label 1 (2) indicates that $\psi_+(x)$ is even (odd). The transmission amplitude for a plane wave incident from one side is found by taking a suitable superposition of these even/odd parity states, giving 
$t(\epsilon,q)=\frac12(z_1-z_2)$. In this way we find
\be\label{eq:delta_function}
t(\epsilon,q)
=\frac12\lp \frac{\lambda_{-\kappa}+ir\lambda_{-k}}{\lambda_{-\kappa}-ir\lambda_k}
-\frac{\lambda_k+ir\lambda_\kappa}{\lambda_{-k}-ir\lambda_\kappa}
\rp
,
\ee
where $\lambda_{\pm\kappa}=\lambda\pm 2\kappa$, $\lambda_{\pm k}=\lambda\pm 2ik$, $r=q^2/k\kappa$. At small $q$, the first term exhibits a resonance associated with the confined state, with the energy $\epsilon_0$,
whereas the second term is nonresonant. As shown in Fig.\ref{fig1}b, 
the cloak effect suppresses 
 the transmission resonance $\epsilon\approx\epsilon_0$ 
at near-normal incidence angles. 
Away from resonance, transmission as a function of $\epsilon$ and $q$ closely mimics that for the Schr\"odinger problem, Eq.(\ref{eq:coupleEq0}). This is illustrated in  Fig.\ref{fig1}c inset. 

A similar behavior was found for a wide barrier, taken to be  
$U(x)=-\frac12 m\omega^2x^2$.
The transmission probability, obtained  numerically, is shown in Fig.\ref{fig2}a. For normal incidence, this potential supports a family of chirality-induced bound states with evenly-spaced energies $\epsilon_n=-\hbar \omega(n+1/2)$. Accordingly, our simulation yields resonances in transmission and conductance peaked near these values (Figs.\ref{fig2},\ref{fig:G3}). The cloak effect is manifested in resonance widths vanishing at normal incidence.

Since the BLG Hamiltonian supports coexisting evanescent and plane wave states, the procedure for this solution is notably more complicated than that for the non-relativistic Schr\"{o}dinger equation. In particular, Eq.(\ref{eq:coupleEq}) must be solved {\it twice} for each $q$, once with an outgoing plane wave as the initial condition, and once with a decaying evanescent wave.  
The physical solution is the 
linear combination of these two solutions in which the coefficients of the growing evanescent waves on both sides are zero. We evaluate the conductance using the relation 
\be \label{eq:G_general} 
G(\epsilon) = G_0W\int dq|t(\epsilon, q)|^2 
,\quad G_0=\frac{N e^2}{(2\pi)^2 \hbar}, 
\ee 
where $W$ is the width of the sample in the $y$-direction, and $N = 4$ is the spin/valley degeneracy of BLG.  
The integral in 
Eq.(\ref{eq:G_general}) runs over the interval $-k_F<q<k_F$, where $k_F$ is the Fermi wavevector in the leads.

Unlike the case of a delta function barrier, where resonance appeared as a weak feature on top of a large background transmission, transmission at zero $q$ is fairly low for the parabolic barrier. This is so because in the cloaking regime transmission is dominated by direct tunneling without coupling to confined states. Since tunneling is exponentially small for a wide barier, transmission is dominated by resonances. The reduction of direct tunneling compared to resonant transmission is clearly seen in simulation (see Fig.\ref{fig2}b inset).
The nearly flat (black) line is the background transmission, calculated for $q = 0$, which is due to direct tunneling under the barrier.  
The asymmetric lineshape for nonzero $q$, which arise due to interference of the direct and resonant transmission pathways, can be described by the Fano model\cite{Fano},
%
\be \label{eq:Fano} 
T_{\rm Fano}(\epsilon) = \frac{1}{1+a^2}\,\frac{(\epsilon - \epsilon_* + a\gamma)^2}{(\epsilon - \epsilon_*)^2 + \gamma^2}, 
\ee 
where  $\gamma$ and $a$ are the energy width and `asymmetry parameter' that depend on the specifics of the system.  

Higher order resonances 
display more symmetric (approximately Lorentzian) profiles which can be explained by large values of the parameter $a$ in Eq.~(\ref{eq:Fano}). Large $a$ values indeed follow from an estimate based on the general relation derived in Ref.\cite{Nockel}, $a = \left|r_0/t_0\right|$, where $r_0$ and $t_0$ are the background reflection and transmission amplitudes in the absence of the resonance. For a wide barrier, the problem can be described by the WKB approach, giving transmission which is exponentially small, and thus a large $a$ value. In particular, for a parabolic barrier we have $|r_0/t_0|=\exp\lp \pi\epsilon/ \hbar \omega\rp $. Hence for values of $\epsilon$ above a few times $\hbar \omega$, the transmission amplitude near resonance is well-approximated by the Breit-Wigner model,  
\be \label{eq:trans} 
t(\epsilon, q)=\frac{\gamma(q)}{\gamma(q)+i(\epsilon-\epsilon_*(q))}, 
\ee 
where the parameters $\epsilon_*(q)$ and $\gamma(q)$ describe the resonance energy and width as a function of wavevector $q$.
 
What is the origin of the peculiar peak shapes seen in conductance? To elucidate their relation with Lorentzian peaks in transmission, Eq.(\ref{eq:trans}), we examine the quantity $\delta G(\epsilon)=G(\epsilon)-G(\epsilon_*(0))$. Using Eq.(\ref{eq:G_general}), we write 
\be \label{eq:t-t0} 
\delta G(\epsilon)=G_0 W \int \lp |t(\epsilon,q)|^2-|t(\epsilon_*(0),q)|^2 \rp dq
. 
\ee 
Since the difference of the two terms under the integral quickly goes to zero for $\gamma(q)\gg |\epsilon-\epsilon_*(0)|$, near the resonance the integral is dominated by small $q$. 

Using quadratic model dependencies $\epsilon_*(q)=\epsilon_*(0)+\alpha q^2$, $\gamma(q)=\beta q^2$ which are valid at small $q$, we can simplify the expression for $\delta G$ as follows 
\be \nonumber 
\delta G(\epsilon) =\frac{G_0 W \beta^{3/2}}{\beta^2+\alpha^2}\int_{-\infty}^\infty 
du \frac{2\tilde \alpha \delta\epsilon u^2 -\delta\epsilon^2}{u^4 
+(\delta\epsilon-\tilde\alpha u^2)^2}
,\quad
\tilde\alpha=\frac{\alpha}{\beta}
.
\ee 
where $\delta\epsilon=\epsilon-\epsilon_*(0)$, $u=\sqrt{\beta}q$. The parameter $\tilde\alpha$ provides a relative measure of the rates at which the resonance moves and widens as a function of $q$. Integration can be done using the partial fraction decomposition, 
\be 
\frac{2\tilde \alpha u^2 \delta\epsilon -\delta\epsilon^2}{u^4 +(\delta\epsilon-\tilde\alpha 
u^2)^2}=\frac{(\tilde\alpha+i)\delta\epsilon}{u^2+i(\tilde\alpha u^2-\delta\epsilon)}+ 
\mbox{c.c.}
, 
\ee 
and the identity $\int_{-\infty}^\infty \frac{du}{au^2+b}=\pi(ab)^{-1/2}$, where for complex $a$ and $b$ the branch of the square root corresponding to ${\rm Re}\,\sqrt{b/a}>0$ should be used. This gives 
\be \label{eq:G} \delta G(\epsilon)=-G_0 W\frac{\cos\lp 
\frac{3\phi}{2} + \frac{\pi}{4}\mbox{sgn}\,(\delta\epsilon) \rp}{(1+\tilde\alpha^2)^{\frac{3}{4}}}\sqrt{\frac{|\delta\epsilon|}{\beta}}, 
\ee 
%
$\phi = \arctan(\tilde\alpha)$. This analysis predicts a square root singularity near the tips,
$\delta G(\epsilon)\propto \sqrt{|\delta\epsilon|}$. The values $\alpha$ and $\beta$ are determined by the details of the barrier potential, and lead to four possible singular lineshapes \cite{online}. Our numerical results confirm this analysis: peaks in transmission are approximately Lorentzian (see Fig.~\ref{fig2} inset), whereas the conductance peaks
feature sharp cusps approximately described by $\delta G(\epsilon) \propto \sqrt{|\delta\epsilon|}$, see Figs.\,\ref{fig2},\ref{fig:G3}.

As discussed above, the cloaking of confined states becomes imperfect in the presence of magnetic field, since the cloaking condition $q=0$ is violated by trajectory bending. 
This effect can be incorporated in our analysis via a momentum shift $q\to q+(eB/c\hbar)x$, giving  $\delta q=(eB/c\hbar)L$ for a change in momentum of a particle moving across the barrier, where $L$ is the barrier width. The effect of a weak field can be mimiced
by introducing a cutoff $\delta q$ in the integral (\ref{eq:t-t0}), leading to rounding the square-root cusps on the scale $\delta\epsilon\sim B^2L^2$. Thus, uncloaking of confined states in weak fields will manifest itself in suppression and rounding of 
conductance peaks.

It is straightforward to incorporate the effect of interlayer gap opening in the analysis \cite{online}.
As shown in Fig.\ref{fig:G3}, the opening of a gap weakly modifies the positions and shapes of the resonances, {\it but does not destroy them}. 
In particular, the fringe contrast, as assessed by the max/min ratio, is relatively unaffected by the gap (see Fig.\ref{fig:G3} inset). 
Therefore the resonances in conductance provide a robust signature of confined states. The peak shapes are singular when these states are cloaked, and smeared when they are uncloaked by applying a magnetic field or an interlayer gap. These features provide a characteristic signature of cloaking and uncloaking, which we expect to be readily observable in experiments. 

The phenomena described above originate from decoupling of chiralities for normally incident states leading to formation of confined states within the barrier. Because such behavior is diametrically opposite to that studied in single layer graphene, it might be called `anti-Klein tunneling'. However, since the term `Klein tunneling' is in a sense a misnomer describing a non-tunneling behavior, we prefer to avoid introducing distinctions based on the degree of kleinness. 

To summarize, chirality mismatch of states inside and outside a ballistic p-n-p junction results in cloaking of states confined inside the junction. 
The cloak effect is perfect at normal incidence angle, and close to perfect at near-normal incidence.  
The confined states manifest themselves as resonances in the ballistic conductance with characteristic non-Lorentzian lineshapes with square-root singularities due to cloaking. These singular resonances, which are smeared by weak magnetic fields,
can serve as a hallmark of the cloak effects in transport measurements.  
 
This work was supported by Office of Naval Research Grant No. N00014-09-1-0724.

\section{Appendix A: singular shapes of the conductance peaks}

\begin{figure} 
\includegraphics[width=3.0in]{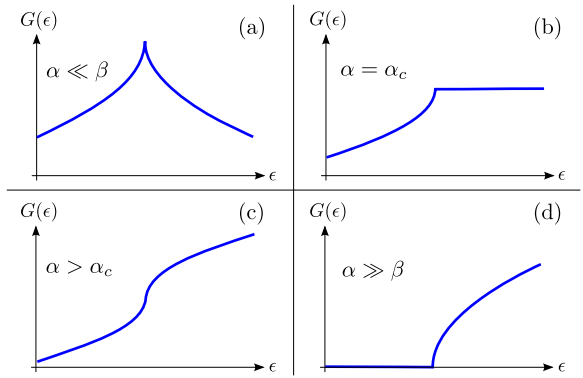} 
\caption[]{Schematic for the conductance peak shape $G(\epsilon)$ 
for different values of the parameter $\alpha$ describing the dispersion of the quasibound state, $\epsilon_*(q)=\epsilon_*+\alpha q^2$, Eq.(\ref{eq:G}). Small $\alpha\ll\beta$ corresponds to a cusp with a square-root profile (a). For $\alpha$ near the critical value $\alpha_c = \beta/\sqrt{3}$ the peak flattens out on one side (b). 
For $\alpha > \alpha_c$, the peak transforms into a monotonic lineshape (c), (d). 
The results for positive $\alpha$ (shown) and for negative $\alpha$ are related by symmetry $\delta\epsilon\to-\delta\epsilon$. 
} 
\vspace{-1mm}
\label{fig:lineshapes} 
\end{figure}

As shown in the main text (see Eq.(\ref{eq:G}) and accompanying discussion), the shape of conductance peaks is sensitive to the relative rates at which the transmission resonances shift and widen as a function of the parallel momentum $q$. The different types of behavior, illustrated in Fig.~\ref{fig:lineshapes}, can be understood as follows. In the limit $\alpha \ll \beta$, the position of the resonance is approximately stationary as a function of $q$, producing a symmetric lineshape (since the transmission profile is symmetric). 
In this case, $\delta G(\epsilon)$ is negative, forming a cusp at $\epsilon = \epsilon_*$ [Fig.~\ref{fig:lineshapes}(a)]. 
As $\alpha$ increases, the contributions of transmission with different $q$ values move to one side of the resonance, 
resulting in a conductance lineshape that is asymmetric about $\epsilon_*$. For 
a sufficiently rapidly shifting resonance, at the critical value $\alpha_c = \beta/\sqrt{3}$, the leading term for $\delta G(\epsilon)$ on one side of the resonance vanishes, giving a flattened behavior $\delta G$ {\it vs.} $\epsilon$ [Fig.~\ref{fig:lineshapes}(b)]. Increasing $\alpha$ above $\alpha_c$ changes the sign of $\delta G(\epsilon)$ [Fig.~\ref{fig:lineshapes}(c)]. For $\alpha \gg \beta$, the resonance position moves fast as a function of $q$ resulting in nonzero $G(\epsilon)$ on only one side of the resonance [Fig.~\ref{fig:lineshapes}(d)]. 

\begin{figure} 
\includegraphics[width=3.25in]{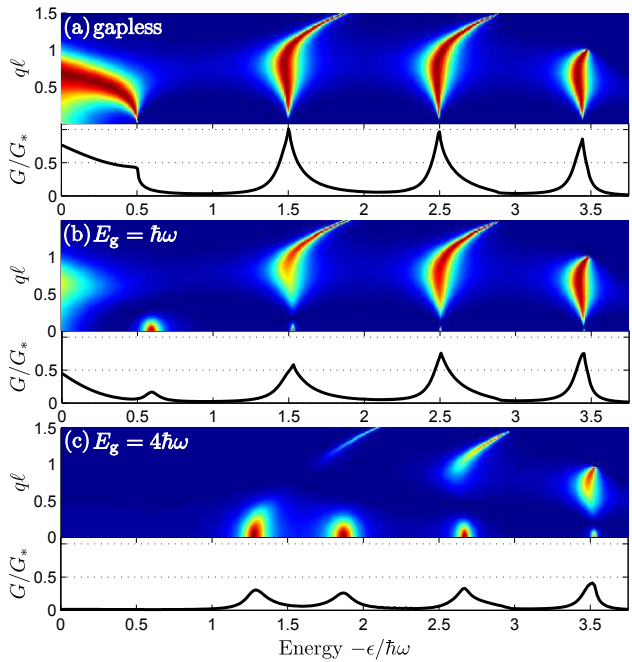} 
\caption[]{Transmission and conductance of the gapless system (top), and gapped systems with $E_{\rm g} = \hbar \omega$ (middle) and $E_{\rm g} = 4 \hbar \omega$ (bottom). Axes labels are the same in Fig.~\ref{fig2}. Shown are resonances correspond to the  harmonic oscillator states $\epsilon = -\frac12 \hbar \omega$, $-\frac32 \hbar \omega$, $-\frac52 \hbar \omega$, $-\frac72 \hbar \omega$. As the gap $E_{\rm g}$ increases, it starts to dominate the resonant tunneling coupling, producing a rounding of the suare-root singularities at the peak tips.}
\vspace{-4mm}
\label{fig:GT1357} 
\end{figure} 

\section{Appendix B: The interlayer gap effect}
The generalization of our problem to BLG in the presence of non-zero transverse polarization that opens up an interlayer gap $E_{\rm g}$ is described by adding a term $\frac{1}{2}E_{\rm g}\sigma_3$ to the Hamiltonian (\ref{eq:H}), $H_{\rm gap}=H+\frac{1}{2}E_{\rm g}\sigma_3$. Writing the Schrodinger equation in pseudospin components $\sigma_x=\pm 1$, we obtain a generalization of Eq.~(\ref{eq:coupleEq}):
%
\be \label{eq:coupleEq_gap} 
\left[-\frac{d^2}{dx^2} - q^2 \pm \frac{2m}{\hbar^2}(U(x)-\epsilon)\right]\psi_\pm =  
\left[2q\frac{d}{dx} \mp \frac{mE_{\rm g}}{\hbar^2}\right]\psi_\mp, 
\ee 
From the coupled equations Eq.(\ref{eq:coupleEq_gap}), we see that for $E_{\rm g} \neq 0$ and at $q = 0$, there is a coupling between the confined and deconfined states $\psi_+$ and $\psi_-$ of order $E_{\rm g}$. This coupling limits the width of a resonance in transmission by $\gamma(q)\propto E_{\rm g}^2$ when $q = 0$, which sets the minimum cusp width for conductance peaks to be of order $E_{\rm g}^2$ (see Fig.~\ref{fig:GT1357}). We also note that there is a line of transmission zeros away from $q = 0$, which corresponds to the situation where the right hand side of Eq.(\ref{eq:coupleEq}) is zero. If we approximate $(\frac{d}{dx})^{-1}$ as $\lambda = \hbar/\sqrt{2m\epsilon}$, we find a curve described by $q\ell = \frac{E_{\rm g}/\hbar \omega}{4\sqrt{\epsilon/\hbar \omega}}$, which shows that the transmission zero gets closer to $q = 0$ for increasing $\epsilon$. Also, the curve collapses to $q = 0$ when $E_{\rm g}$ goes to zero, as expected from the cloak effect.

\end{document}